\documentclass[12pt,preprint,titlepage]{aastex}

\newcommand{\CIVdblt}{{C}\kern 0.1em{\sc iv}~$\lambda\lambda 1548, 1550$}
\newcommand{\MgIIdblt}{{Mg}\kern 0.1em{\sc ii}~$\lambda\lambda 2796, 2803$}
\newcommand{\SiIVdblt}{{Si}\kern 0.1em{\sc iv}~$\lambda\lambda 1393, 1402$}
\newcommand{\NVdblt}{\hbox{{N}\kern 0.1em{\sc v}~$\lambda\lambda 1239,1243$}}
\newcommand{\OVIdblt}{{O}\kern 0.1em{\sc vi}~$\lambda\lambda 1032, 1038$}
\newcommand{\AlIIIdblt}{{Al}\kern 0.1em{\sc iii}~$\lambda\lambda 1855, 1863$}
\newcommand{\CII}{\hbox{{C}\kern 0.1em{\sc ii}}}
\newcommand{\CIII}{\hbox{{C}\kern 0.1em{\sc iii}}}
\newcommand{\CIV}{\hbox{{C}\kern 0.1em{\sc iv}}}
\newcommand{\HI}{\hbox{{H}\kern 0.1em{\sc i}}}
\newcommand{\NaI}{\hbox{{Na}\kern 0.1em{\sc i}}}
\newcommand{\HII}{\hbox{{H}\kern 0.1em{\sc ii}}}
\newcommand{\HeI}{\hbox{{He}\kern 0.1em{\sc i}}}
\newcommand{\HeII}{\hbox{{He}\kern 0.1em{\sc ii}}}
\newcommand{\AlII}{\hbox{{Al}\kern 0.1em{\sc ii}}}
\newcommand{\AlIII}{\hbox{{Al}\kern 0.1em{\sc iii}}}
\newcommand{\NII}{\hbox{{N}\kern 0.1em{\sc ii}}}
\newcommand{\Lya}{\hbox{{Ly}\kern 0.1em$\alpha$}}
\newcommand{\Lyb}{\hbox{{Ly}\kern 0.1em$\beta$}}
\newcommand{\Lyg}{\hbox{{Ly}\kern 0.1em$\gamma$}}
\newcommand{\Lyd}{\hbox{{Ly}\kern 0.1em$\delta$}}
\newcommand{\Lye}{\hbox{{Ly}\kern 0.1em$\epsilon$}}
\newcommand{\FeII}{\hbox{{Fe}\kern 0.1em{\sc ii}}}
\newcommand{\MgI}{\hbox{{Mg}\kern 0.1em{\sc i}}}
\newcommand{\MgII}{\hbox{{Mg}\kern 0.1em{\sc ii}}}
\newcommand{\OVI}{\hbox{{O}\kern 0.1em{\sc vi}}}
\newcommand{\OVII}{\hbox{{O}\kern 0.1em{\sc vii}}}
\newcommand{\OVIII}{\hbox{{O}\kern 0.1em{\sc viii}}}
\newcommand{\NV}{\hbox{{N}\kern 0.1em{\sc v}}}
\newcommand{\SiII}{\hbox{{Si}\kern 0.1em{\sc ii}}}
\newcommand{\SiIII}{\hbox{{Si}\kern 0.1em{\sc iii}}}
\newcommand{\SiIV}{\hbox{{Si}\kern 0.1em{\sc iv}}}
\newcommand{\kms}{\ensuremath{\mathrm{km~s^{-1}}}}
\newcommand{\cmsq}{\ensuremath{\mathrm{cm^{-2}}}}
\newcommand{\cc}{\ensuremath{\mathrm{cm^{-3}}}}
\newcommand{\Ang}{\hbox{\textrm{\AA}}}

\newcommand{\Zsun}{\ensuremath{Z_{\Sun}}}

\usepackage{epsfig}

\begin{document}

\title{A survey of weak \MgII~absorbers at redshift $\langle z \rangle = 1.78$\footnotemark[1]}

\footnotetext[1]{Based on public data obtained from the ESO archive of observations from the UVES spectrograph at the VLT, Paranal, Chile.}

\author{Ryan S. Lynch\altaffilmark{2}, Jane C. Charlton\altaffilmark{2}, and Tae-Sun Kim\altaffilmark{3,4}}

\altaffiltext{2}{Department of Astronomy and Astrophysics, The Pennsylvania State University, University Park, PA 16802, \textit{rlynch, charlton@astro.psu.edu}}
\altaffiltext{3}{Institute of Astronomy, Madingley Road, Cambridge}
\altaffiltext{4}{Astrophysikalisches Institut Potsdam, An der Sternwarte 16, 14482 Potsdam, Germany, \textit{tkim@aip.de}}

\begin{abstract}

The exact nature of weak \MgII~absorbers (those with $W_r(2796) < 0.3~\Ang$) is a matter of debate, but most are likely related to areas of local star formation or supernovae activity outside of giant galaxies.  Using 18 QSO spectra obtained with the Ultra-Violet Echelle Spectrograph (UVES) on the Very Large Telescope (VLT), we have conducted a survey for weak \MgII~absorbers at $1.4 < z < 2.4$.  We searched a redshift path length of $\Delta z = 8.51$, eliminating regions badly contaminated by atmospheric absorption so that the survey is close to 100\% complete to $W_r(2796) = 0.02~\Ang$.  We found a total of 9 weak absorbers, yielding a number density of absorbers of $dN/dz = 1.06 \pm 0.12$ for $0.02 \leq W_r(2796) < 0.3~\Ang$.  \citet{Nar05} found $dN/dz = 1.00 \pm 0.20$ at $0 < z < 0.3$ and \citet{Church99} found $dN/dz = 1.74 \pm 0.10$ at $0.4 < z < 1.4$.  Therefore, the population of weak \MgII~absorbers appears to peak at $z \sim 1$.  We explore the expected evolution of the absorber population subject to a changing extragalactic background radiation (EBR) from $z = 0.9$ to $z = 1.78$ (the median redshift of our survey), and find that the result is higher than the observed value.  We point out that the peak epoch for weak \MgII\ absorption at $z\sim1$ may coincide with the peak epoch of global star formation in the dwarf galaxy environment.

\end{abstract}


\section{Introduction}
\label{sec:intro}

Quasar absorption line systems (QALS) are unique and powerful tools for studying the chemical content, kinematics, ionization state, and overall structure of galaxies as well as of the intergalactic medium over $0 < z < 6$.  \MgII~absorption is an important feature in many QALS, because it relates to metal forming processes.  The \MgIIdblt~doublet is a strong transition and is easily detected in low ionization QALS, making it an extremely useful tool for probing galaxies and their environments and for studying the structure of the intergalactic medium.  Weak \MgII~absorbers are defined to be those absorbers with $W_r(2796) < 0.3~\Ang$, and at least some of them represent a unique population(s) of \MgII~absorbers that appears to be much different from stronger absorbers \citep{Rig02, Nest05}.

Unlike strong \MgII~absorbers, which are typically associated with luminous galaxies (within $\sim38 h^{-1} (L/L^*)^{0.15}$ kpc) \citep{Berg91, Berg92, LeBrun93, Steid94, Steid97, Steid95}, weak \MgII~absorbers at $0.4 < z < 1.4$ are not typically found within $50 h^{-1}$ kpc of such a galaxy \citep[but see \citet{Church05} for some exceptions]{Rig02}, raising questions about their processes of formation.  Through both statistical means \citep{Church99, Rig02} and through investigation of individual systems \citep{Church00}, weak \MgII~absorbers are found to arise in sub-Lyman limit systems ($15.8 < \log{N(\HI)} < 16.8~[\cmsq]$), comprising at least 25\% of \Lya~forest clouds in that column density range \citep{Rig02}.  Weak \MgII~absorbers are also found to have high metallicity; at least 10\% solar, but in some cases even solar or supersolar \citep{Rig02, Char03,Simcoe05}.  This is a striking result because, as noted above, luminous galaxies are rarely found within a $50 h^{-1}$ kpc impact parameter of the quasar.  Furthermore, some absorbers have \FeII/\MgII~ratios that do not allow for $\alpha$-enhancement solely from Type II supernovae, suggesting that metals are also produced ``\textit{in situ}'' by Type Ia supernovae and not ejected by nearby, luminous galaxies.  It is possible that some weak \MgII~absorbers arise in areas of intergalactic star formation, or are related to supernovae remnants in dwarf galaxies \citep{Rig02}.  Their incidence is therefore likely to be related to global star formation rates and supernovae activity in these environments. 

\citet{Milni05} have found that a flattened geometry for weak \MgII~absorbers is most consistent with their number statistics and kinematics, and suggest that such structures could be part of the cosmic web.  Photoionization modeling suggests that the \MgII~absorption arises in a region $\sim 1-100$ pcs thick \citep{Char03,Simcoe05}.  This thin region that produces \MgII~absorption is most likely surrounded by a lower density region that produces high ionization \CIV~absorption, with both regions centered at the same velocity \citep{Rig02}.  Components offset in velocity by tens of \kms~also often exist in \CIV~and could represent additional surrounding low density filaments \citep{Milni05}.

Previous surveys for weak \MgII~absorbers have been conducted at $0 < z < 0.3$ by \citet{Nar05} and at $0.4 < z < 1.4$ by \citet{Church99}.  \citet{Nar05} used E140M Space Telescope Imaging Spectrograph \textit{Hubble Space Telescope} archive data to search for \SiII~$\lambda 1260$ and \CII~$\lambda 1335$ as tracers of \MgII, and found a number density of absorbers $dN/dz = 1.00 \pm 0.20$.  \citet{Church99} used the High Resolution Spectrograph on the Keck I Telescope and observed the \MgIIdblt~doublet directly, finding $dN/dz = 1.74 \pm 0.10$.  These values imply that the total cross section for weak \MgII~absorption is twice that of strong \MgII~absorbers \citep{Steid92}.

The findings of \citet{Nar05} for $0 < z < 0.3$ weak \MgII~absorbers are consistent with cosmological evolution of the $z \sim 1$ weak \MgII~absorber population, based upon $\Lambda$CDM cosmology \citep{Church99}.  However, when the effect of the evolution of the extragalactic background radiation (EBR) on the absorber population is taken into account, $dN/dz$ is expected to be higher at $z \sim 0$ than observed.  This is partly because the less intense ionizing background (compared to $z \sim 1$) leads to an increase in the low-ionization \MgII~phase, i.e. absorbers that were below the equivalent width threshold at $z \sim 1$ are detectable at $z \sim 0$.  Also, there is a significant contribution at $z \sim 0$ from \MgII~absorption in a lower density phase that gave rise only to higher ionization absorbers at $z \sim 1$ \citep{Nar05}.  It appears that the processes that lead to weak \MgII~absorber formation are less active at $z \sim 0$ then at $z \sim 1$.  Determining the evolution of $dN/dz$ at higher redshifts is an important step toward identifying the formation processes of weak \MgII~absorbers, and thus served as the motivation for this work.

In order to push observational measurements of $dN/dz$ for weak \MgII~absorbers to higher redshifts, we surveyed the spectra of 18 QSOs obtained with the Ultra-Violet Echelle Spectrograph (UVES) on the Very Large Telescope (VLT).  Such a measurement will shed light on the nature of the absorber population, and may indicate which factors affect the evolution of the absorbers.  The data and survey method are outlined in \S~\ref{sec:survey}.  Our survey results are summarized in \S~\ref{sec:results}.  We also simulated the evolution of the $z = 0.9$ absorber population back to $z = 1.78$, subject to the changing EBR in order to predict the expected $dN/dz$ evolution.  These results are presented in \S~\ref{sec:evol} and compared with our survey results.  In \S~\ref{sec:conc} we summarize our findings and end with a discussion.

\section{Data and Survey Method}
\label{sec:survey}

\subsection{VLT Spectra}
\label{sec:spectra}

\begin{figure}
\centering
\vspace{0.0in}
\epsscale{1.0}
\plotone{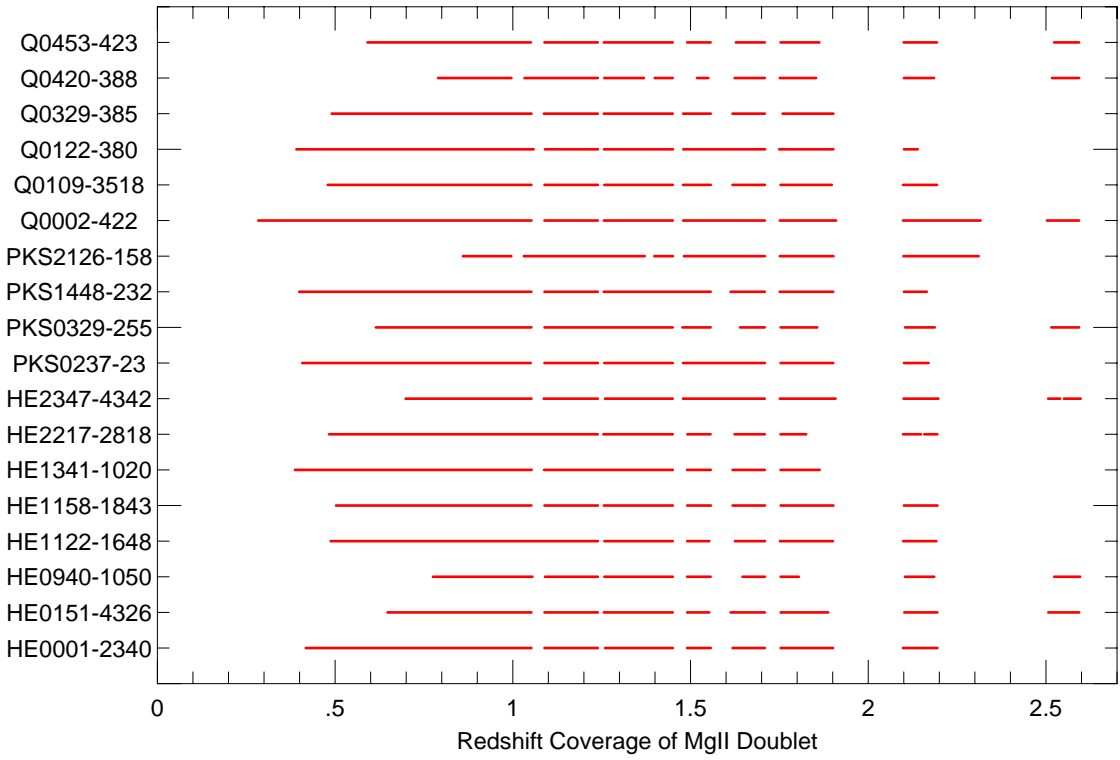}
\caption{The redshift coverage for the \MgIIdblt~doublet in each of our 18 QSO spectra.  Our survey covers a total redshift path length of $\Delta z = 8.51$ for $1.4 < z < 2.4$. \label{fig:redcovplot}}
\end{figure}

The 18 UVES/VLT QSO spectra used for our survey of weak \MgII~absorption were obtained from the ESO archive.  The spectra typically covered a wavelength range of $\approx 3050 - 10075~\Ang$, though there were usually gaps in coverage from $\approx 5753 - 5841~\Ang$ and $\approx 8520 - 8665~\Ang$ (see Figure \ref{fig:redcovplot}).  The resolution was $R \sim 45,000$, or $\sim 6.7~\kms$.  Data reduction procedures are described in \citet{Kim04}.  Continuum fitting followed standard procedures using the IRAF\footnotemark[4] SFIT task.  A cubic spline function was used, typically of order 7 - 9.

Though the spectra were of extremely high quality, there were some complicating issues.  As already mentioned, there were some small breaks in the coverage in many of the spectra.  The dense \Lya~forest made a $z < 0.7$ \MgIIdblt~doublet survey impractical for some quasars, but this was not a concern since the \citet{Church99} survey also covered this range.  There was also significant contamination from telluric absorption features, making a survey of affected wavelengths impractical, as well.  Our general philosophy was to exclude troublesome regions rather than to risk missing \MgII~doublets in our survey.

\footnotetext[4]{IRAF is distributed by the National Optical Astronomy Observatories which are operated by AURA, Inc., under contract to the National Science Foundation.}

\subsection{Survey Method}
\label{sec:method}

For our survey, we first assumed that every line detected above a $5 \sigma$ significance level was \MgII~$\lambda 2796$.  It was then determined if the corresponding \MgII~$\lambda 2803$ line was present at the expected location with at least a $2.5 \sigma$ significance level.  If this was the case, the line profiles were checked for similarity in shape by visual inspection, and for consistency with a 2:1 optical depth ratio.  In all cases some supporting lines at the redshift of our \MgII~candidate were found (such as \Lya, \FeII, \AlIII, and/or \CIV).  Once the detection of a \MgII~absorber was confirmed, we measured the rest frame equivalent width.  In order to be accepted as a weak absorber, a rest-frame equivalent width of $0.02 \leq W_r(2796) < 0.3~\Ang$ was required.  We note that our survey could have extended to a much lower equivalent width limit.  The $W_r(2796) = 0.02~\Ang$ limit was chosen for comparison to previous surveys.  This corresponds to $\log{N(\MgII)} = 11.72~[\cmsq]$ for a typical Doppler parameter of $b = 4~\kms$, for an unsaturated line.  Fortunately, above this limit interference by telluric absorption did not affect our survey in our selected wavelength regions (i.e. telluric features were weaker than $W_r(2796) = 0.02~\Ang$).

For the purpose of the present survey, we focused on absorbers whose redshift fell in the range $1.4 < z < 2.4$.  Several quasars had coverage at $z \sim 2.5$, but the path length was not significant at that $z$.  The large wavelength coverage of these spectra also allowed us to search for weak \MgII~absorbers to significantly lower redshifts in most spectra.  Therefore, we also considered absorbers with redshifts in the range $0.4 < z < 1.4$ for comparison with the results of \citet{Church99}.

The excellent S/N of these spectra allowed our survey to be nearly 100\% complete if we eliminated wavelength ranges with heavy atmospheric absorption.  The telluric features in the eliminated regions acted as ``effective noise'' which made unambiguous detection of a $0.02 \leq W_r(2796) < 0.3~\Ang$ \MgII~$\lambda 2796$ line at a $5\sigma$ significance level difficult, if not impossible.  Figure \ref{fig:redcovplot} shows the redshift coverage for the \MgIIdblt~doublet in each of our spectra.  Because we excluded certain redshifts from our search and because not every QSO spectrum covered the \MgIIdblt~doublet up to $z = 2.4$, our average redshift does not coincide with the center of our redshift range.  Considering these factors, the average redshift is given by
\begin{eqnarray}
\langle z \rangle = \sum\limits_{included~regions}{\frac{\int\limits^{z_{high}}_{z_{low}}{zdz}}{\int\limits^{z_{high}}_{z_{low}}{dz}}} \label{eq:zave}
\end{eqnarray}

In order to verify that we have nearly 100\% completeness in the spectral regions that we surveyed, we randomly added two artificial \MgII~doublets to each of our spectra, in the redshift range $0.4 < z < 2.4$.  Simulated lines were not placed in regions of the spectra that were not searched, i.e. the \Lya~forest, telluric absorption regions, or coverage gaps.  The artificial lines were all assigned the minimum equivalent width for our survey, $W_r(2796) = 0.02~\Ang$.  Lines were added, beginning with the observed flux and convolving it with the line spread function for the simulated line.  This allowed us to retain a realistic representation of noise in the region, making artificial lines indistinguishable from real ones.  The spectra were searched for these artificial lines in the same manner that we searched for real lines.  All the artificial lines were found even though they were at the minimum equivalent width for our survey.  Therefore, we are confident that we effectively have 100\% completeness in the regions we searched.

\section{Results of the Survey}
\label{sec:results}

Survey completeness as a function of redshift was calculated according to the formalism given in \citet{Steid92} and \citet{Lanz87}.  The total redshift path covered in the 18 QSO spectra over the range $1.4 < z < 2.4$ is given by
\begin{eqnarray}
Z(W_r,R) = \int\limits_{1.4}^{2.4} g(W_r, z, R) dz \label{eq:zpath}
\end{eqnarray}
where $g(W_r, z, R)$ is the function that gives the number of lines of sight along which the \MgIIdblt~doublet could have been detected at a redshift \textit{z} and with a rest frame equivalent width $W_r(2796)$ at greater than or equal to a $5 \sigma$ significance level.  \textit{R} is the expected ratio of equivalent widths between \MgII~$\lambda 2796$ and \MgII~$\lambda 2803$, which, according to atomic physics, is two for unsaturated lines.  As described in \S~\ref{sec:method}, we find 100\% completeness for all \MgII~doublets with $W_r(2796) > 0.02~\Ang$.  Therefore, we have the simplest case in which $Z(0.02~\Ang,2)= \Delta z = 8.51$, the full path length covered by our survey.

\begin{figure}
\centering
\vspace{0.0in}
\epsscale{0.3}
\plotone{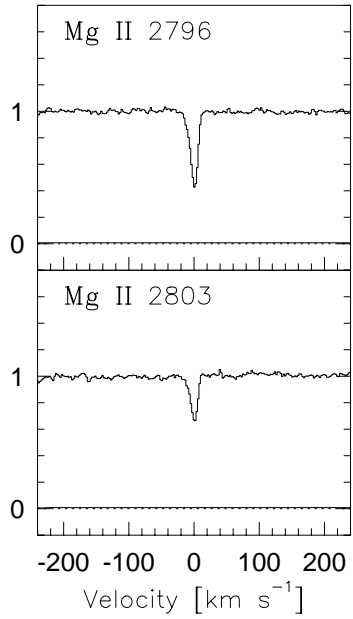}
\caption{Absorption profiles, plotted in velocity space, for \MgIIdblt~for the $W_r(2796) = 0.076 \pm 0.001~\Ang$ absorber at $z = 1.405362$ toward HE2347-4342.   \CIVdblt~is also detected at this redshift. \label{fig:HE2347a}}
\end{figure}

\begin{figure}
\centering
\vspace{0.0in}
\epsscale{0.3}
\plotone{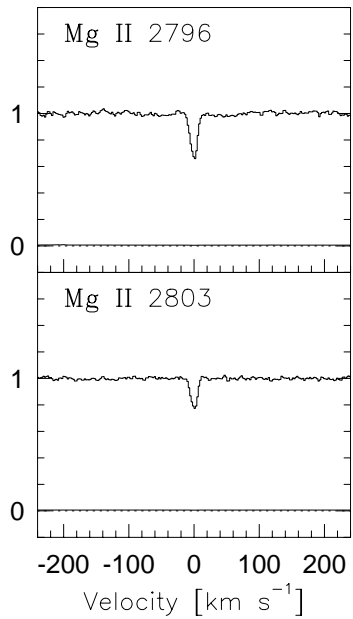}
\caption{The $W_r(2796) = 0.042 \pm 0.001~\Ang$ absorber at $z = 1.446496$ toward Q0002-422.  \CIVdblt~is also detected at this redshift. \label{fig:Q0002}}
\end{figure}

\begin{figure}
\centering
\vspace{0.0in}
\epsscale{0.3}
\plotone{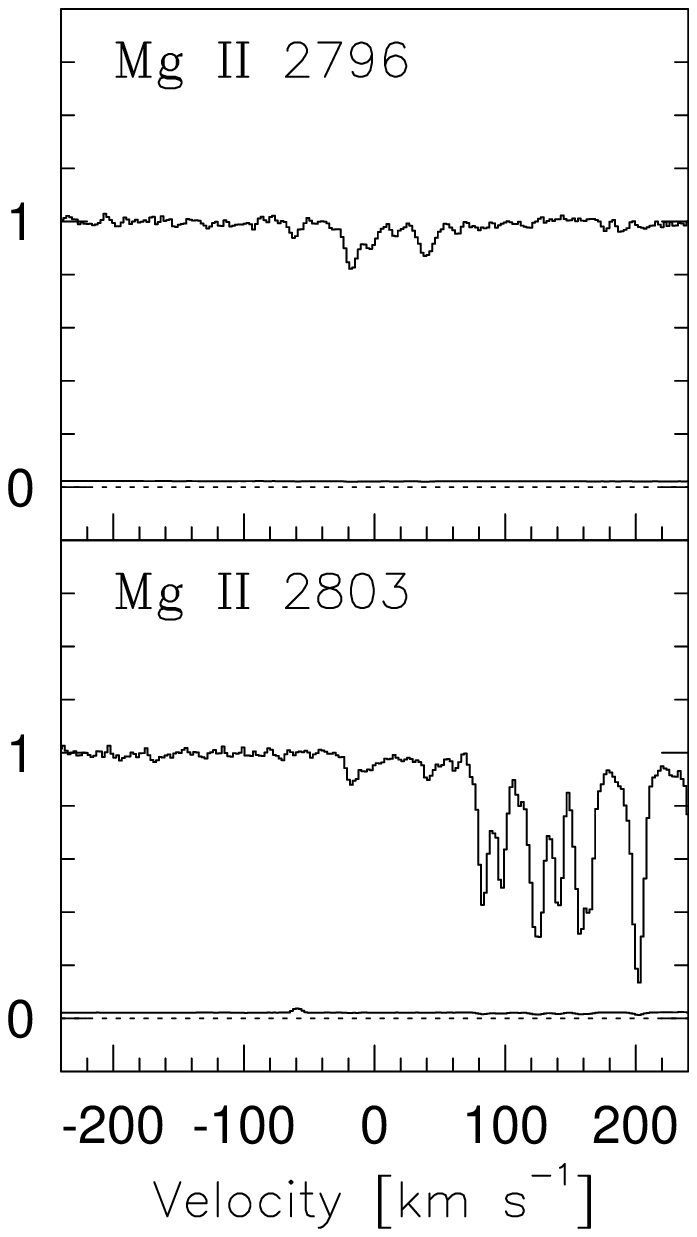}
\caption{The $W_r(2796) = 0.061 \pm 0.003~\Ang$ absorber at $z = 1.450109$ toward Q0122-380.  \SiIVdblt~and \CIVdblt~are also detected at this redshift. \label{fig:Q0122}}
\end{figure}

\begin{figure}
\centering
\vspace{0.0in}
\epsscale{0.3}
\plotone{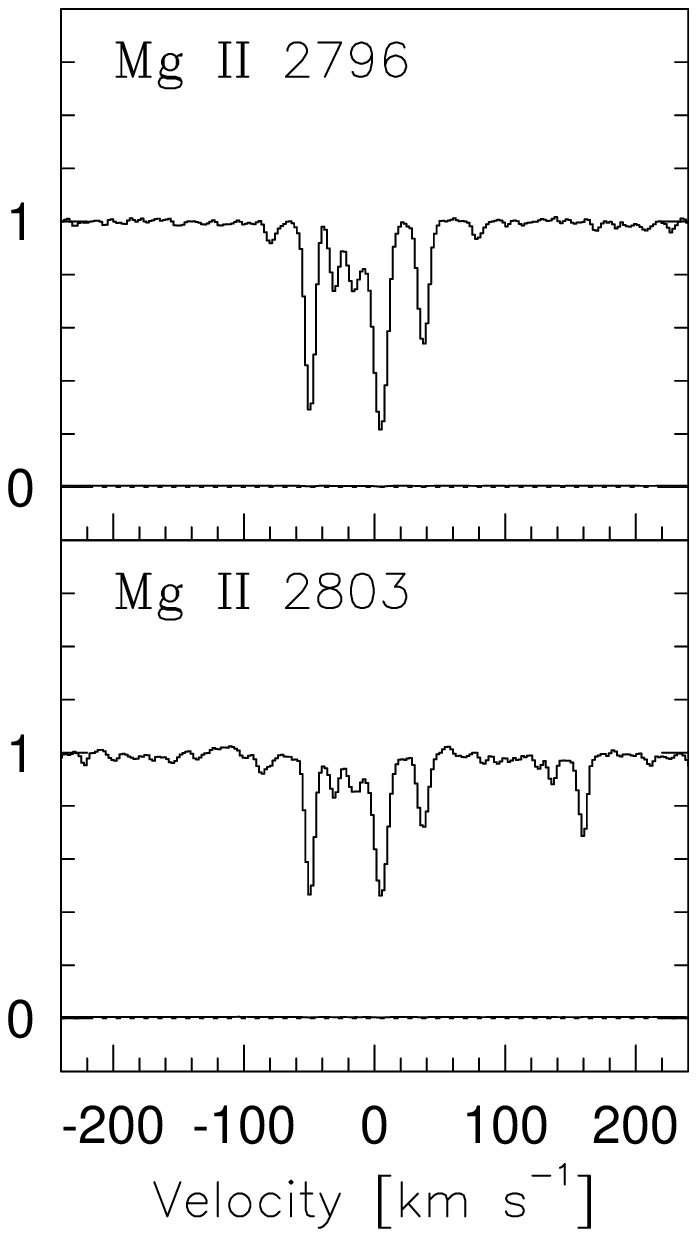}
\caption{The $W_r(2796) = 0.268 \pm 0.001~\Ang$ absorber at $z = 1.555845$ toward HE2217-2818.  \Lya, \SiII~$\lambda 1260$, \CII~$\lambda 1335$, \SiIVdblt, \CIVdblt, \AlII~$\lambda 1671$, \FeII~$\lambda 2344$, \FeII~$\lambda 2383$, and \FeII~$\lambda 2600$ are also detected at this redshift. \label{fig:HE22171}}
\end{figure}

\begin{figure}
\centering
\vspace{0.0in}
\epsscale{0.3}
\plotone{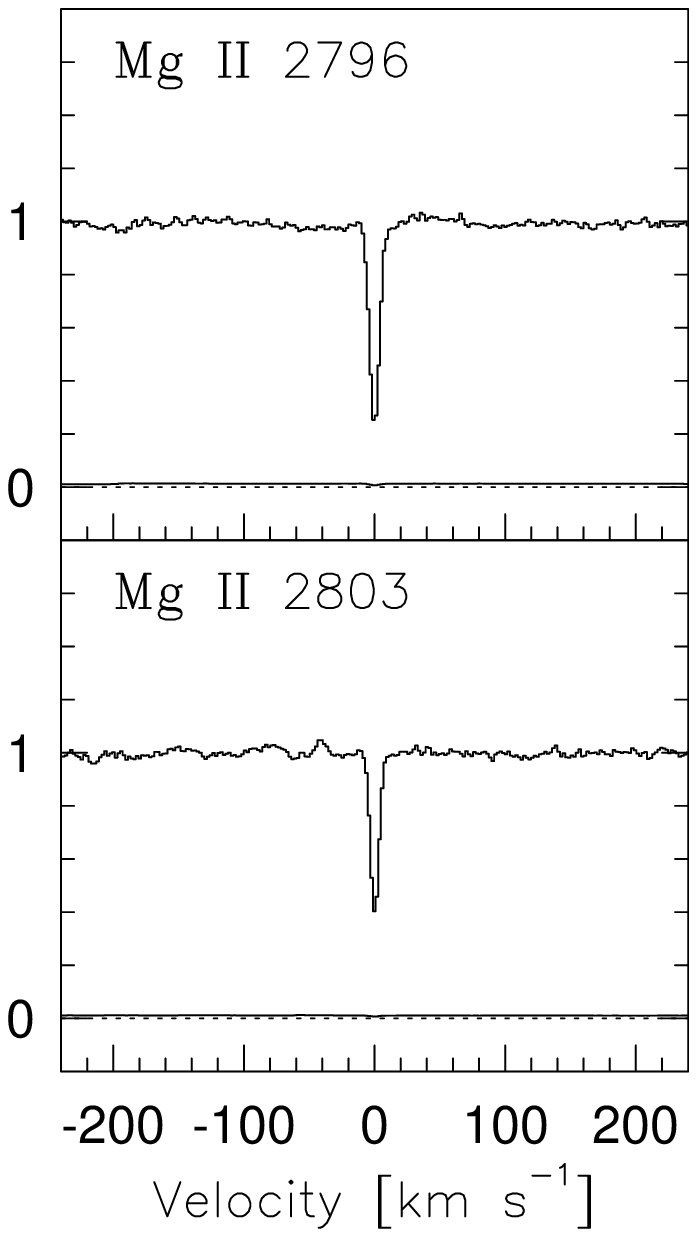}
\caption{The $W_r(2796) = 0.070 \pm 0.001~\Ang$ absorber at $z = 1.651462$ toward HE0001-2340.  \Lya, \SiIII~$\lambda 1207$, \SiII~$\lambda 1260$, \CII~$\lambda 1335$, \SiIVdblt, \SiII~$\lambda 1527$, and \CIVdblt, are also detected at this redshift. \label{fig:HE0001}}
\end{figure}

\begin{figure}
\centering
\vspace{0.0in}
\epsscale{0.3}
\plotone{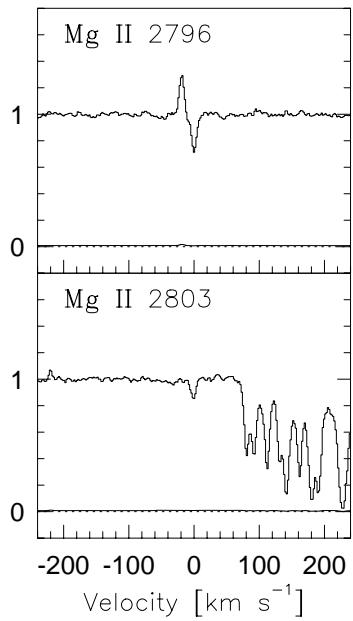}
\caption{The $W_r(2796) = 0.027 \pm 0.001~\Ang$ absorber at $z = 1.708494$ toward HE0151-4326.  \Lya, \SiIVdblt, and \AlIIIdblt~are also detected at this redshift.  The expected locations of many other common features fell in the \Lya-forest, making their identification difficult.  The anomaly near the \MgII~$\lambda 2796$ line is a residual due to poor sky subtraction. \label{fig:HE0151}}
\end{figure}

\begin{figure}
\centering
\vspace{0.0in}
\epsscale{0.3}
\plotone{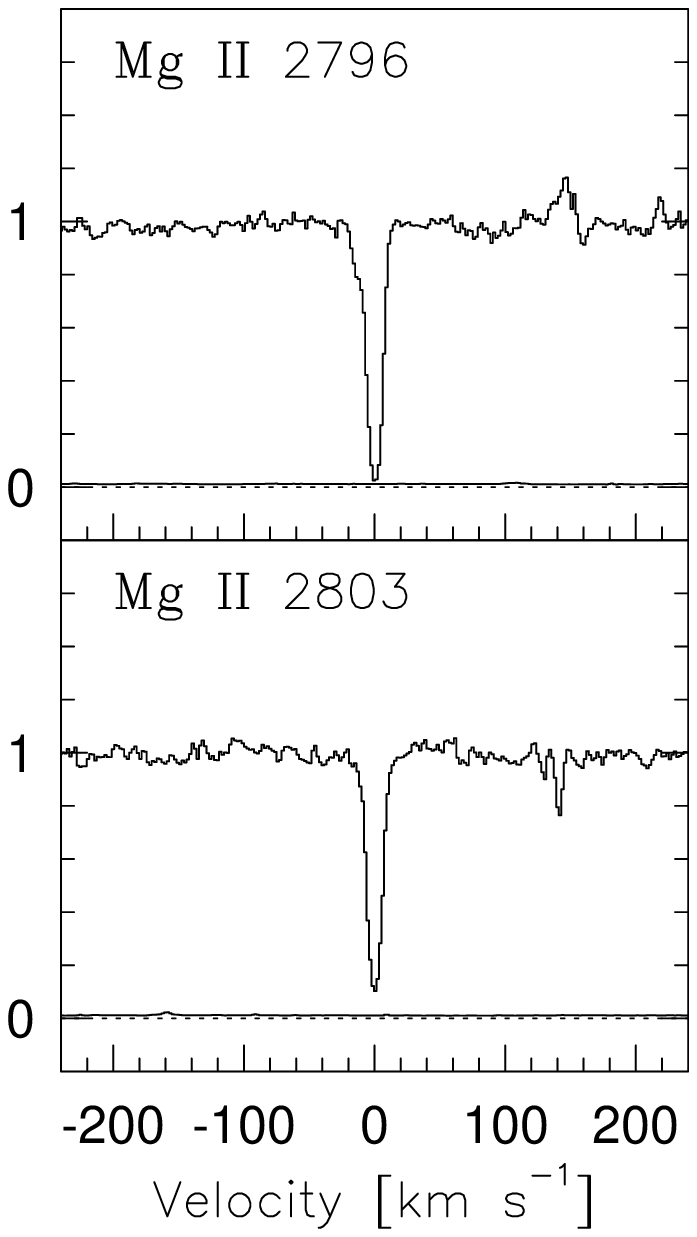}
\caption{The $W_r(2796) = 0.146 \pm 0.001~\Ang$ absorber at $z = 1.796237$ toward HE2347-4342.  \Lya, \SiII~$\lambda 1260$, \CII~$\lambda 1335$, \SiIVdblt, \SiII~$\lambda 1527$, \CIVdblt, \AlIIIdblt, \FeII~$\lambda 2383$, and \FeII~$\lambda 2600$ are also detected at this redshift. \label{fig:HE2347b}}
\end{figure}

\begin{figure}
\centering
\vspace{0.0in}
\epsscale{0.3}
\plotone{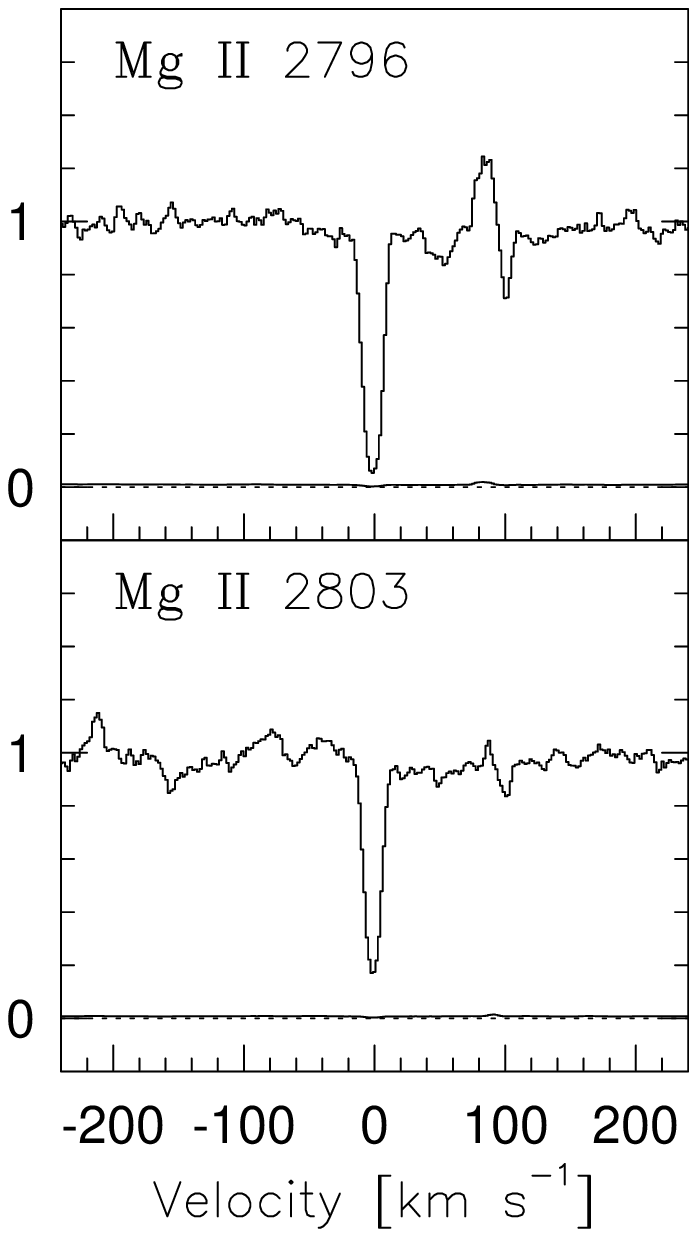}
\caption{The $W_r(2796) = 0.254 \pm 0.002~\Ang$ absorber at $z = 1.858380$ toward Q0453-423.  \Lya, \SiII~$\lambda 1190$, \SiII~$\lambda 1193$, \SiIII~$\lambda 1207$, \SiII~$\lambda 1260$, \CII~$\lambda 1335$, \AlIII~$\lambda 1671$, \AlIIIdblt, \FeII~$\lambda 2344$, \FeII~$\lambda 2374$, \FeII~$\lambda 2383$, and \FeII~$\lambda 2600$ are also detected at this redshift. \label{fig:Q0453}}
\end{figure}

\begin{figure}
\centering
\vspace{0.0in}
\epsscale{0.3}
\plotone{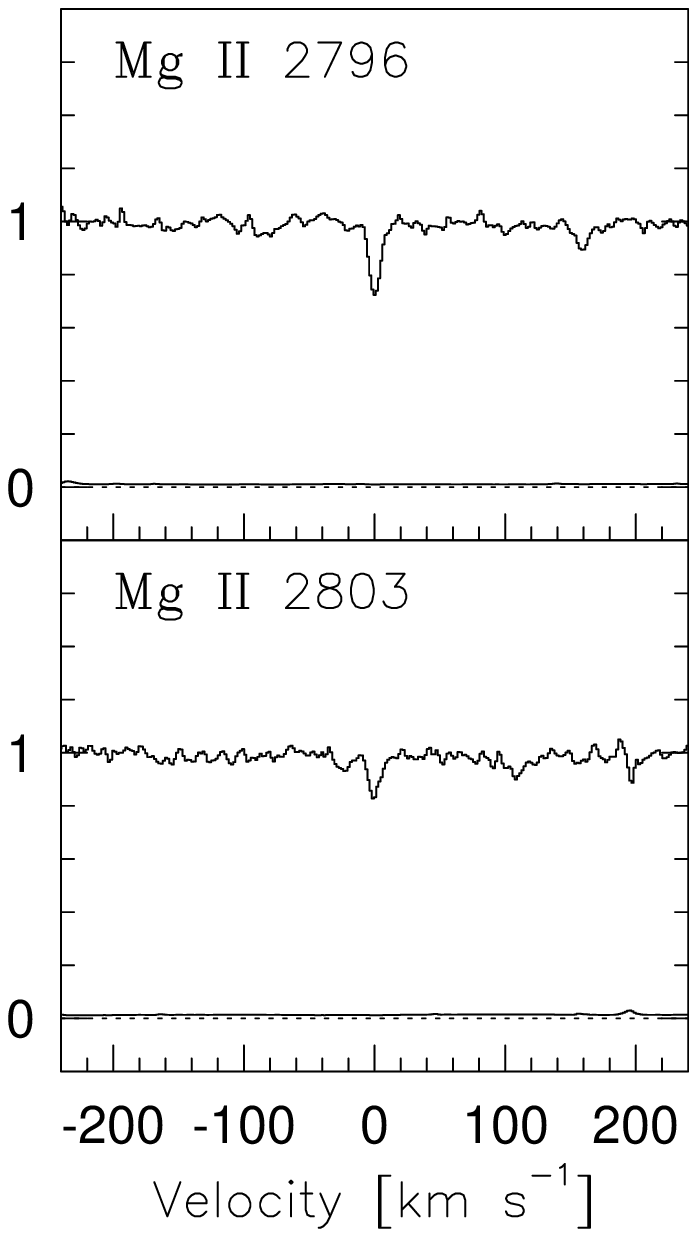}
\caption{The $W_r(2796) = 0.028 \pm 0.001~\Ang$ absorber at $z = 2.174546$ toward HE0940-1050.  \Lya, \SiIII~$\lambda 1207$, \SiII~$\lambda 1260$, \CII~$\lambda 1335$, \SiIVdblt, and \CIVdblt~are also detected at this redshift. \label{fig:HE0940}}
\end{figure}

\begin{figure}
\centering
\vspace{0.0in}
\epsscale{1.0}
\plotone{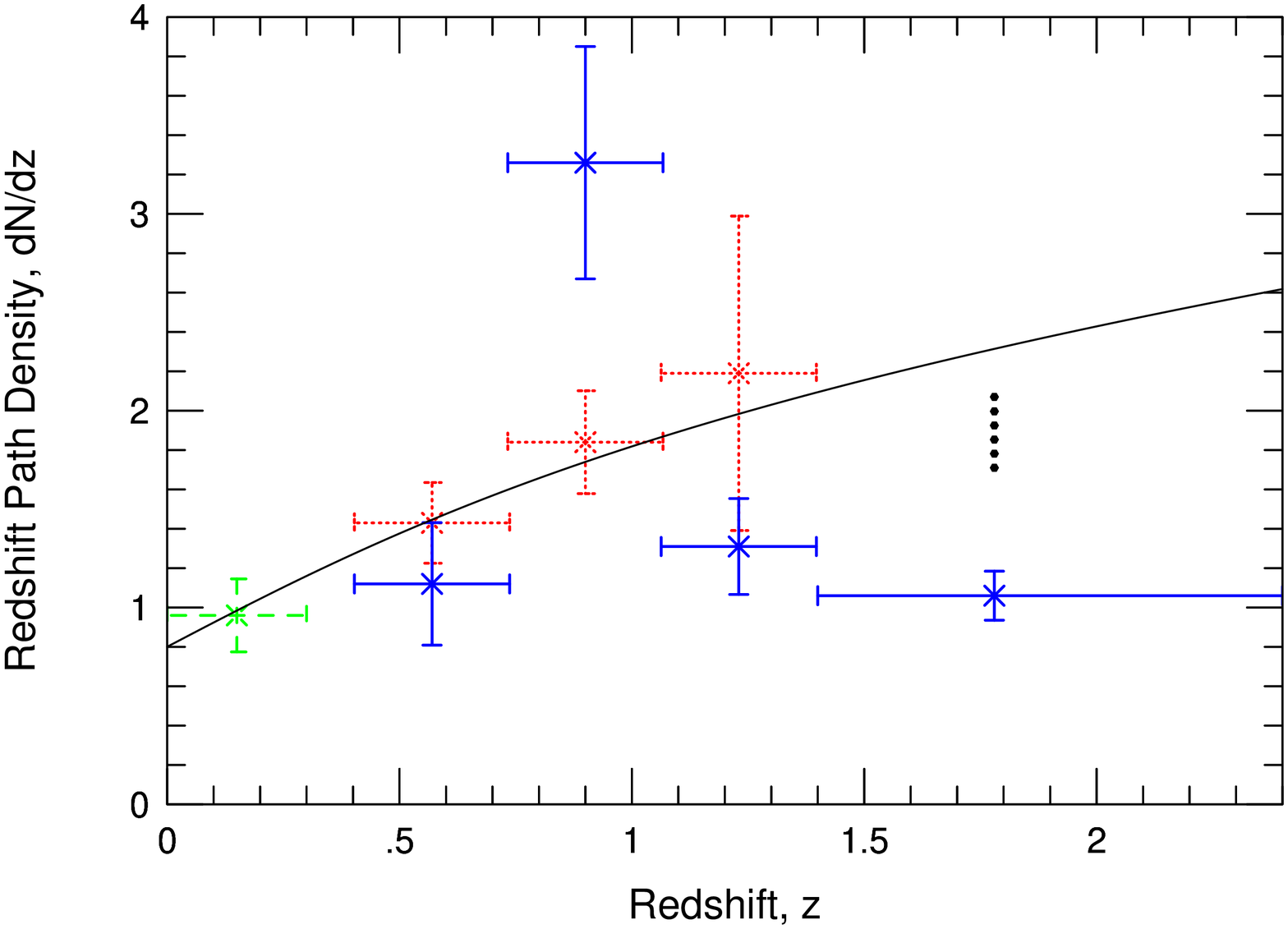}
\caption{The redshift path density ($dN/dz$) for our survey over the redshift range $1.4 < z < 2.4$ is presented as a point with solid error bars.  We find $dN/dz = 1.06 \pm 0.12$.  We also present our results for the redshift range $0.4 < z < 1.4$ with solid error bars, for comparison to the results of \citet{Church99}, that are shown with dotted error bars.  The result of \citet{Nar05} is shown with dashed error bars.  The solid surve shows the expected evolution of absorbers based an a $\Lambda$CDM cosmology.  The series of black dots indicates the range of expected $dN/dz$ for our survey based upon the evolving EBR and a $\Lambda$CDM cosmology, assuming a static population. \label{fig:dndzplot}}
\end{figure}

For the $1.4 < z < 2.4$ \MgII~doublets found in our survey, Table \ref{table:sysprop} lists the absorber redshifts (calculated from the optical depth mean) and the quasar in whose spectrum they were detected.  Column 3 lists the velocity width of the systems (as defined in \citet{Church01}), Column 4 the rest frame equivalent width of \MgII~$\lambda 2796$, and Column 5 the \MgII~doublet ratio, $DR = W_r(2796)/W_r(2803)$.  Their \MgII~$\lambda 2796$ and \MgII~$\lambda 2803$ profiles are shown in Figures \ref{fig:HE2347a}-\ref{fig:HE0940}.  We find 9 systems over a redshift path of $\Delta z = 8.51$ at $1.4 <z < 2.4$ in the 18 spectra surveyed.  
Because our survey is 100\% complete to $W_r(2796)=0.02~\Ang$, we can simply divide the number of systems by the redshift path to obtain a redshift number density $dN/dz = 1.06 \pm 0.12$.  Figure \ref{fig:dndzplot} shows our result, along with the results of \citet{Nar05} and \citet{Church99}.  We also show our survey results for the redshift range $0.4 < z < 1.4$, separated into the same bins as in \citet{Church99}, for which we find agreement to within $1-2 \sigma$.  Overall, for $0.4 < z < 1.4$, we found 29 systems in a redshift path of $\Delta z = 14.5$, yielding $dN/dz = 2.01 \pm 0.14$.  \citet{Church99} found $dN/dz = 1.74 \pm 0.10$ for a somewhat larger redshift path of $\Delta z = 17.2$.  The solid curve shows the no-evolution expectation for $dN/dz$ in a $\Lambda$CDM universe ($\Omega_m = 0.3$ and $\Omega_{\Lambda} = 0.7$) normalized to $dN/dz = 1.74$ at $ z= 0.9$.

\section{Expected Evolution of the Weak \MgII~Absorber Population Due to the Changing EBR}
\label{sec:evol}

The number density of weak \MgII~absorbers changes with redshift for a variety of reasons.  Since we expect weak \MgII~absorption to arise in small-scale regions, the process of hierarchical structure growth could destroy the absorbers through mergers or collapse.  As the metallicity of the universe increases, we expect to see an increase in the number of metal-rich systems.  Also, the cumulative effects of supernovae, causing superbubbles and superwinds, will redistribute gas in the universe.  The ionization state of weak \MgII~absorbers is also expected to change as the EBR evolves.  The EBR is known to change with redshift because of the change in the number density of bright quasars and in the rate of global star formation \citep{Haardt96, Haardt01}.  This changing EBR will have an effect, even on an otherwise static population of absorbers, because of their changing ionization state.  We explore this effect in the following sections.

\subsection{Expected Change in the Ionization State of \MgII~Absorbers}
\label{sec:ionstatechange}

Models by \citet{Rig02} and \citet{Char03} suggest that weak \MgII~absorbers consist of a high density region of \MgII~and a lower density \CIV~phase.  Typical densities are on the order of 0.1 \cc~for the region that gives rise to the majority of the \MgII~absorption and 0.001 \cc~for the \CIV~region.  We conduct a thought experiment to explore how these absorbers evolve.  Let us assume that a given absorber is stable over time, i.e. there is no gravitational collapse, fragmentation, mergers, evaporation, or pressure changes.  In other words, assume that the absorber is subject to no changes except for the evolving EBR.  The total hydrogen column density, $\log{N(\mathrm{H}_{tot})}$, the electron number density, $n_e$, and the metallicity, $\log{Z/\Zsun}$, are all constant.  One may ask how the ionization state of the absorber will change as a result of the changing EBR, between $z = 0.9$ and $z = 1.78$.  At $z = 1.78$ the EBR is  $\approx 0.4$ dex more intense than at $z = 0.9$, though the shape of the spectrum is similar.  This increase in the photon number density, $n_{\gamma}$ leads to a change in the ionization parameter, $U = n_{\gamma}/n_e$.  For a given cloud we can fix $\log{N(\mathrm{H}_{tot})}$ and calculate the change in $\log{N(\HI)}$ and $\log{N(\MgII)}$.  At higher redshift, we find that the increased EBR causes the equivalent width to be larger for the high ionization state \CIV~and smaller for the low ionization state \MgII, compared to intermediate redshifts.

\subsection{Simulated Evolution of Three Systems}
\label{sec:evolsim}

\begin{figure}
\centering
\vspace{0.0in}
\epsscale{0.82}
\plotone{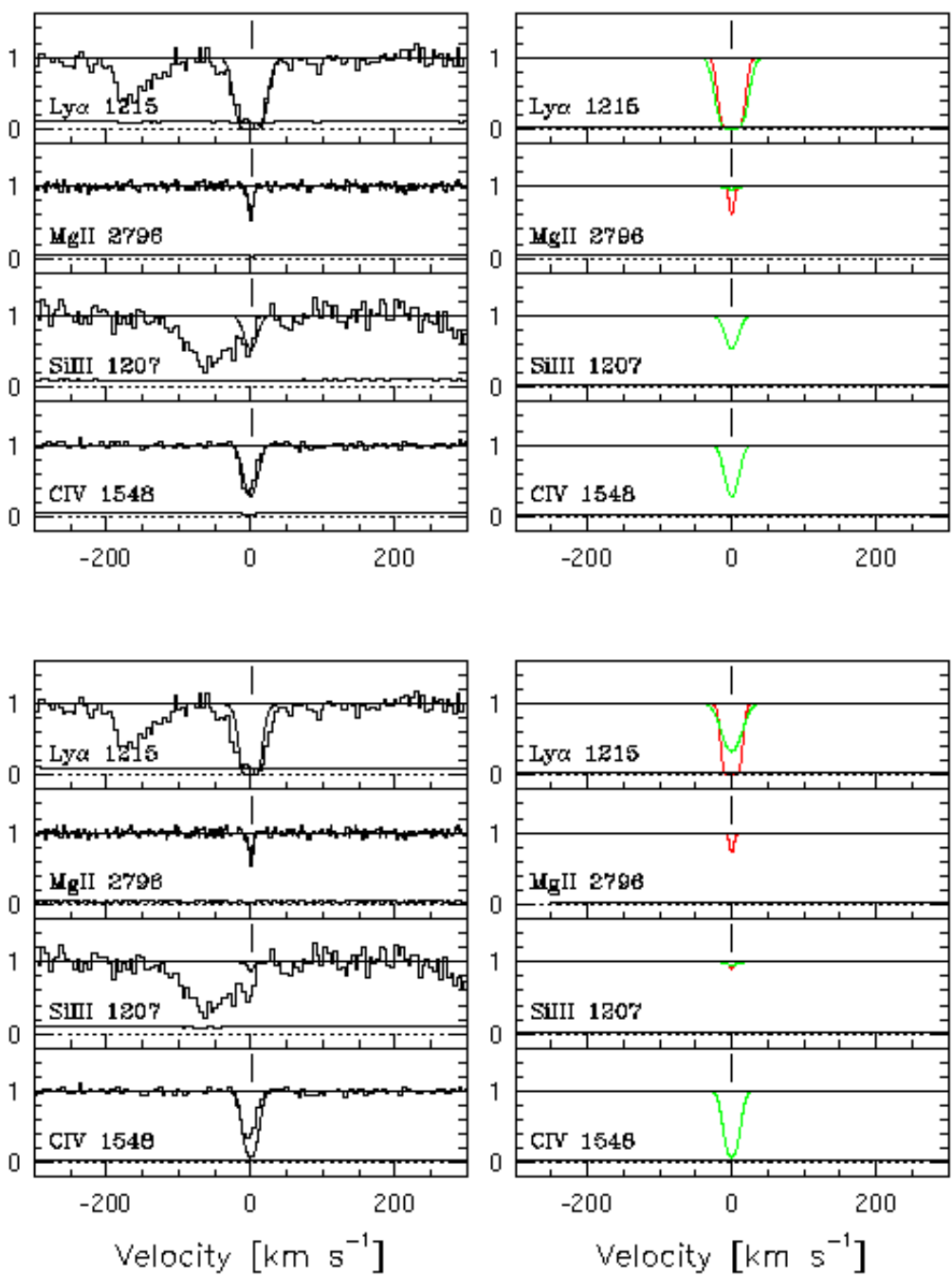}
\caption{The $z = 0.8$ system towards PG1634+706 \citep{Church99}.  The bottom panels show the same system evolved to $z = 1.78$.  The left panels show the actual spectra superimposed with the full models, and the right panels show the various components from the models.  This system has a single cloud in \MgII~(centered at 0~\kms) and, although its single \CIV~component arises in a separate phase, it is aligned with the \MgII~in velocity. \label{fig:sim1}}
\end{figure}

\begin{figure}
\centering
\vspace{0.0in}
\epsscale{0.82}
\plotone{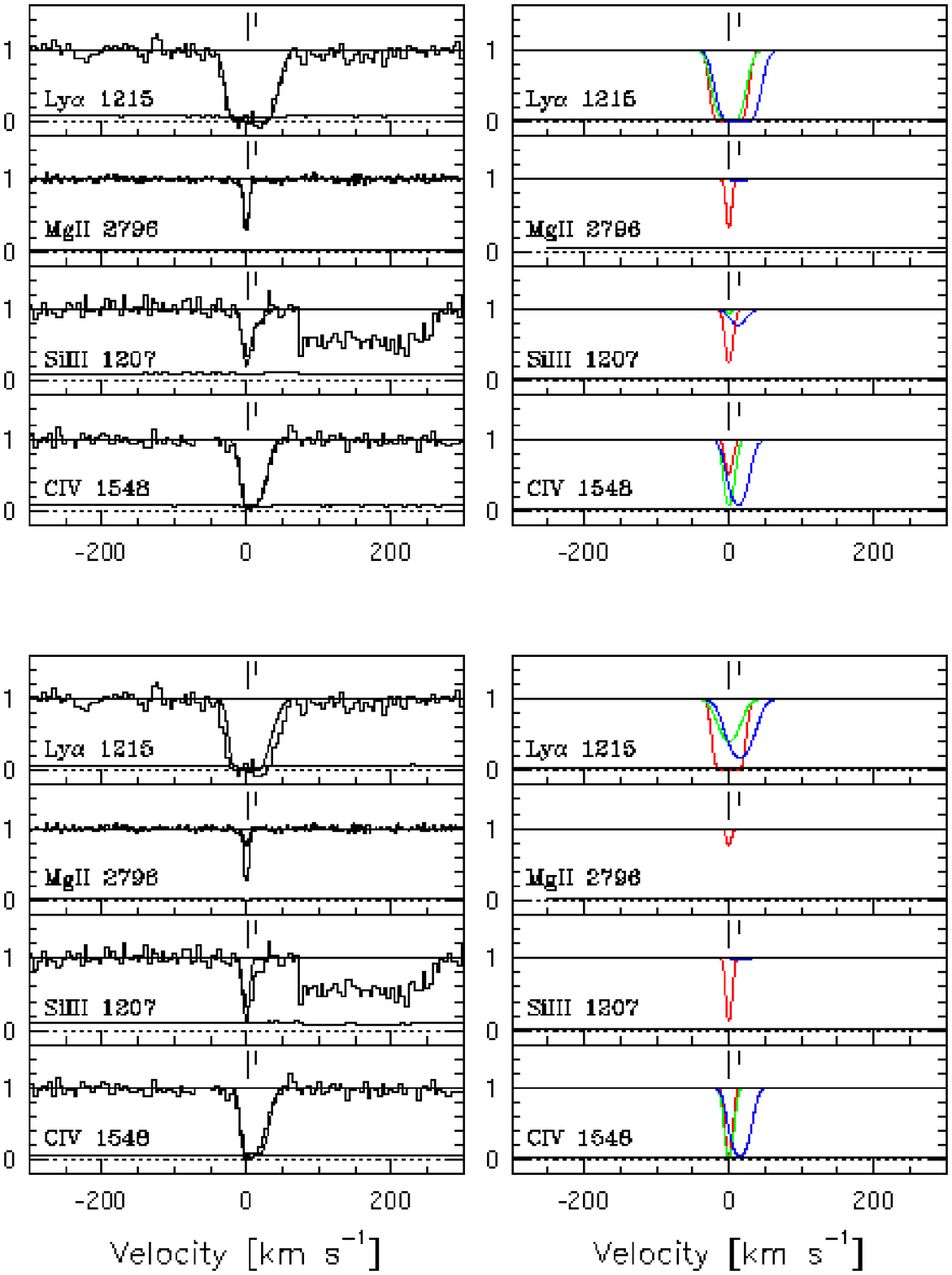}
\caption{The $z = 0.9$ system towards PG1634+706 \citep{Church99}.  The bottom panels show the same system evolved to $z = 1.78$.  The left panels show the actual spectra superimposed with the full models, and the right panels show the various components from the models.  This system has a single cloud in \MgII~(centered at 0~\kms) but has multiple \CIV~components.  \label{fig:sim2}}
\end{figure}

\begin{figure}
\centering
\vspace{0.0in}
\epsscale{0.82}
\plotone{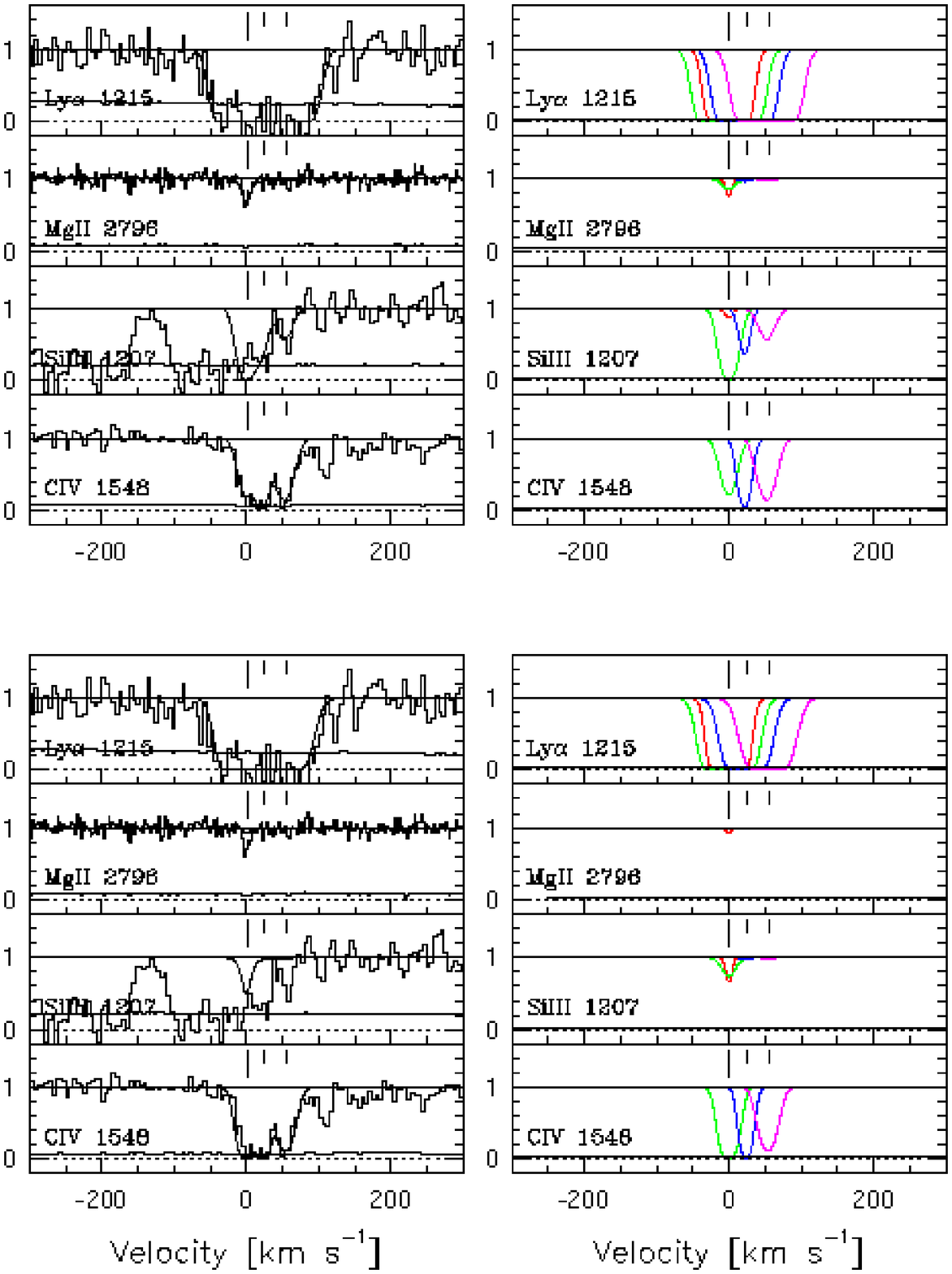}
\caption{The $z = 0.6$ system towards PG1634+706 \citep{Church99}.  The bottom panels show the same system evolved to $z = 1.78$.  The left panels show the actual spectra superimposed with the full models, and the right panels show the various components from the models.  This system has a single component in \MgII~(centered at 0~\kms), but has multiple components in \CIV.  The \MgII~for this system evolves to be below the detection criterion, while the \CIV~becomes stronger.  \label{fig:sim3}}
\end{figure}

To illustrate the evolution of weak \MgII~absorbers, we simulate the appearance at $z = 1.78$ of three systems modeled by \citet{Char03} at $\langle z \rangle = 0.9$ subject to the changing EBR and cosmological evolution.  We applied photoionization models using the code Cloudy (version 94.00; \citealt{Fer98}) with an EBR of the form prescribed by \citet{Haardt96,Haardt01}.  We adopted their model including quasars and star forming galaxies, with an escape fraction of 10\% and including absorption by the intergalactic medium.  The results are presented in Figures \ref{fig:sim1} - \ref{fig:sim3}.  Each of the three systems has only a single cloud detected in \MgII, but their \CIV~absorption profiles differ in strength, kinematic spread, and number of clouds.  The $z \sim 0.9$ (Figure \ref{fig:sim2}) and $z \sim 0.6$ (Figure \ref{fig:sim3}) systems have offset \CIV~components in addition to those centered on the \MgII.

Each system was evolved to $z = 1.78$ subject to the changing EBR.  Once the column density of each component was found, we created a synthetic, noise-free spectrum.  For the $z \sim 0.6$ system, the \MgII~profile evolves below the equivalent width cutoff for our survey (see Figure~\ref{fig:sim3}).  The overall trend for all three systems is for weak \MgII~absorbers to be less prevalent at $z = 1.78$ than at $z= 0.9$, while \CIV~absorption, which traces the high ionization state, becomes stronger.

\subsection{Expected $dN/dz$ at $z = 1.78$ Due to the Changing EBR}
\label{sec:expectdndz}

We apply a simple physical model to predict $dN/dz$ at $z = 1.78$ using the results of \citet{Church99} as a referemce point.  This model only takes the evolving EBR into account but is useful for exploring which factors affect the evolution of $dN/dz$ for weak \MgII~absorbers.  Based on the previous measurement of $dN/dz$ at $z = 0.9$ by \citet{Church99}, we now consider what $dN/dz$ would be expected at $z = 1.78$ subject to the changing EBR.  We first determine the $W_r(2796)$ of a $z = 0.9$ absorber whose ionization state has increased just enough from $z = 0.9$ to $z = 1.78$ that it falls below the detection threshold of our survey.  We compute the column density of \MgII~for an absorber with the threshold $W_r(2796) = 0.02~\Ang$ to be $\log{N(\MgII)} = 11.7~[\cmsq]$, assuming a typical Doppler parameter, $b = 4~\kms$.  Models of weak {\MgII} absorbers \citep{Char03,Rig02} have yielded a range of metallicities, $-1.0 \lesssim \log{Z/\Zsun} \lesssim 0.0$, and a range of densities $10^{-2} \lesssim n_e \lesssim 10^{-1}$~\cc.  To illustrate our method we first choose the values that yield the smallest predicted $dN/dz$ value at $z=1.78$, $\log{Z/\Zsun} = -1.0$ and $n_e = 10^{-2}~\cc$, and solve for the total column density $\log{N(\mathrm{H})_{tot}} = 18.3~[\cmsq]$ subject to the Haardt-Madau EBR at $z = 1.78$, again using Cloudy.  We then compute the $W_r(2796)$ of an absorber with this same total column density subject to the EBR at $z = 0.9$.  We find $W_r(2796) = 0.041~\Ang$.  This is the lowest equivalent width \MgII~absorber at $z = 0.9$ that would have been observed as a weak \MgII~absorber at $z = 1.78$.  A similar procedure was followed to find an upper limit on $W_r(2796)$ for absorbers that exceed the 0.3 \Ang~threshold for weak absorption at $z = 0.9$ but were weak at $z = 1.78$.  Such an absorber will have a total column density $\log{N(\mathrm{H})_{tot}} = 21.4~[\cmsq]$, which will produce a $W_r(2796) = 0.31~\Ang$ absorption at $z = 0.9$.

Once an upper and lower limit on equivalent width at $z = 0.9$ have been obtained, we may integrate over the equivalent width distribution function at this redshift to estimate the number of absorbers per unit redshift that we would expect to observe at $z = 1.78$.  The equivalent width distribution follows a power law form:
\begin{eqnarray}
n(W) dW = C W^{-\delta} dW \label{eq:ewdist}
\end{eqnarray}
where $C \approx 0.4$ and $\delta = 1.04$ \citep{Church99}.  The expected ratio of $dN/dz$ at $z = 1.78$ to $dN/dz$ at $z = 0.9$, due only to the changing EBR, is given by
\begin{eqnarray}
\frac{(dN/dz)_{1.78}}{(dN/dz)_{0.9}} = \frac{\int\limits_{0.041}^{0.31}{n(W) \label{eq:dndzratio}
 dW}}{\int\limits_{0.02}^{0.3}{n(W) dW}} = 0.74
\end{eqnarray}
Cosmological evolution of the absorber population must also be taken into account.  For a model with $\Omega_{\Lambda} = 0.7$ and $\Omega_m = 0.3$, cosmological evolution alone would predict a factor of 1.33 more absorbers at $z = 1.78$.  Thus, when we combine evolution due to the EBR and cosmology, we find a predicted $dN/dz = 1.71$ at $z = 1.78$.

If we vary the input cloud properties, such that $\log{Z/\Zsun} = 0.0$ and $n_e = 10^{-1}~\cc$, we find a predicted $dN/dz = 2.14$ at $z=1.78$.  In that case, the electron density change dominates, and the metallicity change only has a small effect.  For this estimate we assumed that typical weak \MgII\ doublets are resolved so that $b=4~\kms$ is a typical value.  There may be some unresolved saturation in some of the lines in our sample, since the doublet ratios of weak lines are not equal to $2$ (see Table \ref{table:sysprop}).  Assuming $b=2~\kms$, we see no significant change in our predicted $dN/dz$.  Also, $b=2~\kms$ implies $T<5800$~K, somewhat smaller than the equilibrium temperature Cloudy finds for our assumed density and metallicity.  An even smaller $b$, which we consider unlikely, would result in somewhat smaller $dN/dz$ predictions, but would also require a very cold cloud.  Thus we believe that $b=4~\kms$ provides a realistic estimate.

The predicted $dN/dz$ of $1.71-2.14$ is significantly higher than the observed value of $dN/dz=1.06\pm0.12$ for weak \MgII\ absorbers at $\langle z \rangle = 1.78$.  This implies that the population of objects that give rise to weak \MgII\ absorption are becoming more common at $z=0.9$ than they were at $z=1.78$.  This could occur either through gradual formation of these objects, assuming that they are not destroyed, or through an increase in the importance of processes that generate a non-static population.

\section{Summary and Discussion}
\label{sec:conc}

We have conducted a survey of 18 QSO spectra in the ESO archive obtained with the UVES on the VLT, searching for weak \MgII~absorbers with $0.02 \leq W_r(2796) < 0.3~\Ang$.  In order to test our survey completeness we randomly added simulated \MgIIdblt~transitions to the spectra.  These simulated lines were at the low end of the equivalent width limit.  Because the spectra have high signal-to-noise, the survey is nearly 100\% complete in the regions that we searched, having eliminated regions strongly affected by telluric absorption.  We found 9 systems in a redshift path length of $\Delta z = 8.51$ over the range $1.4 < z < 2.4$.  This yields a number density per unit redshift of $dN/dz = 1.06 \pm 0.12$ at $\langle z \rangle = 1.78$.  For comparison, \citet{Nar05} find $dN/dz = 1.00 \pm 0.20$ at $\langle z \rangle = 0.15$ and \citet{Church99} find $dN/dz =  1.74 \pm 0.10$ at $\langle z \rangle = 0.9$.  We conclude that the weak \MgII\ absorber population peaks at $z \sim 1$, and that the objects that generate this type of absorption are less common both at higher and at lower redshifts.

A decrease in $dN/dz$ from $z=1.78$ to $z=0.9$ would be expected simply on the basis of cosmological evolution.  The strength of \MgII\ absorption coming from a population of objects is also expected to change subject to the changing EBR. In order to understand the balance of these effects, we calculated the expected $dN/dz$ at $z = 1.78$ based upon the known population at $z = 0.9$.  For the purpose of this estimate, we assumed that the absorber population was static or that destruction was balanced by the formation processes between these two epochs, i.e. we only accounted for evolution due to the changing EBR and due to cosmological evolution.  Our prediction of $dN/dz = 1.71-2.14$ is markedly higher than our observed value.  Therefore, we conclude that one of two general scenarios must apply.  First, the structures that give rise to weak \MgII\ absorption could be long-lived, and there could be a gradual build-up of these structures over time, such as through hierarchical structure formation.  Second, the structures could be transient, such that the rate of their formation process determines $dN/dz$.  An obvious example of such a process is star formation.  

This second possibility was advocated by \citet{Nar05} who considered the evolution of $dN/dz$ from $z=0.9$ to $z=0$.  In that case, starting with the observed $z=0.9$ population of weak \MgII\ absorbers, the evolving EBR along with cosmological evolution would predict a significantly larger population at $z=0$ than is observed.  This issue was complicated by the incidence of a new type of low redshift weak \MgII\ absorber descended from absorbers from which only \CIV\ absorption was observed at $z=0.9$.  Nonetheless, the trend for an overall decrease in the weak \MgII\ absorber population is consistent with the decrease in the global star formation rate over that redshift interval.  This suggests a possible relationship between star formation and weak \MgII\ absorbers, which is also consistent with their high metallicities and (in at least some cases) iron abundances \citep{Rig02}.

We should therefore consider how the global star formation rate relates to the results of our $\langle z \rangle = 1.78$ survey.  The global star formation rate is roughly constant over the redshift range $1 \leq z \leq 4$, and then decreases to lower redshifts (see \citet{Gab04} and references therein).  If production of weak \MgII\ absorbers was directly correlated with the global star formation rate we would expect a $dN/dz$ at $z=1.78$ consistent with a constant creation rate, subject to the EBR and to cosmological evolution.  Our observation of a smaller $dN/dz$ than this expectation would suggest that the global star formation rate should be smaller at $z=1.78$ than at $z=0.9$.  This is not the case.  However, we need to consider that weak \MgII\ absorbers are preferentially found in certain environments.

Theoretically, in the context of hierarchical structure formation and empirically on the basis of the morphology--density relationship, it is expected that star formation peaks later in low mass galaxies \citep{Kauffmann04}.
More directly, \citet{Bauer05} found a shift over the range $0 < z < 1.5$ in the relative contributions of low and high mass galaxies to the global star formation rate, with low mass galaxies contibuting more at low redshift.   Although measurements of the dwarf galaxy contribution have not yet extended to higher redshifts, it seems likely that the star formation rate in dwarfs peaks at a lower redshift than that for giants.  Given the constant global star formation rate at $1 < z < 4$ \citep{Gab04}, if the giants are contributing relatively more at high redshift, the dwarfs must contribute less.  Given previous suggestions that weak \MgII\ absorbers are related to dwarfs, it is intriguing that the peak epoch of weak \MgII\ absorbers may coincide with the peak of star formation in the dwarf environment.

\citet{Rig02} estimate that weak \MgII\ absorbers account for approximately 25\%-100\% of \Lya\ forest clouds with $15.8 < \log N(\HI) < 16.8~[{\rm cm}]^{-2}$ at $z\sim1$.  At first glance it seems surprising that the redshift path density for \Lya\ forest clouds would sharply decrease with decreasing $z$ until $z=1.5$ ($dN/dz \propto (1+z)^{2.47\pm0.18}$) \citep{Kim02,Wey98}, while we find the weak \MgII\ absorber population to increase over the same interval. However, if the weak \MgII\ absorbers are closely related to star formation in dwarfs, as we have suggested, then there is no reason to expect a direct correlation of their evolution with the evolution of the \Lya\ forest, the latter which is explained by cosmological expansion. 

This work was funded by the National Science Foundation grant NSF AST-07138 and by an NSF REU supplement.  We are indebted to the ESO Archive for making this work possible.  The authors wish to express their gratitude to Chris Churchill for his valuable insight and discussions during this project.  Thanks also to Mike Eracleous for his assistance with our analysis.  We also acknowledge Joe Masiero for providing us with tools for our data processing and Andrew Mshar for technical assistance.  We also wish to thank an anonymous referee for invaluable and helpful suggestions.

\begin{deluxetable}{lcrcc}
\setlength{\tabcolsep}{0.15in}
\tablewidth{0pc}
\tablecolumns{5}
\tablecaption{Properties of $1.4 < z < 2.4$ Weak \MgII~Systems \label{table:sysprop}}
\tablehead{\colhead{QSO} & \colhead{$z_{abs}$} & \colhead{$\omega_{v}$} & \colhead{$W_r(2796)$} & \colhead{\textit{DR}} \\
                         &                     & \colhead{({\kms})}     & \colhead{({\AA})}           &}
\startdata
HE2347-4342 & 1.405362 & $5.8 \pm 0.2$  & $0.076 \pm 0.001$ & $1.84 \pm 0.04$ \\
\hline
Q0002-422   & 1.446496 & $4.9 \pm 0.2$  & $0.042 \pm 0.001$ & $1.65 \pm 0.04$ \\
\hline
Q0122-380   & 1.450109 & $28.4 \pm 0.8$ & $0.061 \pm 0.003$ & $1.34 \pm 0.12$ \\
\hline
HE2217-2818 & 1.555845 & $30.9 \pm 0.1$ & $0.268 \pm 0.001$ & $1.44 \pm 0.01$ \\
\hline
HE0001-2340 & 1.651462 & $4.0 \pm 0.3$  & $0.070 \pm 0.001$ & $1.51 \pm 0.03$ \\
\hline
HE0151-4326 & 1.708494 & $4.0 \pm 0.3$  & $0.027 \pm 0.001$ & $1.96 \pm 0.10$ \\
\hline
HE2347-4342 & 1.796237 & $5.3 \pm 0.1$  & $0.146 \pm 0.001$ & $1.19 \pm 0.01$ \\
\hline
Q0453-423   & 1.858380 & $38.6 \pm 0.4$  & $0.254 \pm 0.002$ & $1.21 \pm 0.01$ \\
\hline
HE0940-1050 & 2.174546 & $6.9 \pm 0.4$  & $0.028 \pm 0.001$ & $1.56 \pm 0.10$ \\
\enddata
\end{deluxetable}

\clearpage
\newpage

\end{document}